\documentstyle[preprint,aps,epsfig,floats]{revtex}

\newcommand{\be}{\begin{equation}}
\newcommand{\ee}{\end{equation}}
\newcommand{\bea}{\begin{eqnarray}}
\newcommand{\eea}{\end{eqnarray}}
\newcommand{\bml}{\begin{mathletters}}
\newcommand{\eml}{\end{mathletters}}

\begin{document}

\tighten

\preprint{DCPT-02/71}
\draft




\title{Interaction of global and local monopoles}
\renewcommand{\thefootnote}{\fnsymbol{footnote}}

\author{Eug\^enio R. Bezerra de Mello\footnote{emello@fisica.ufpb.br}}
\address{Departamento de F\'{\i}sica-CCEN, Universidade Federal da 
Para\'{\i}ba, 58.059-970, J. Pessoa, PB, C. Postal 5.008, Brazil}
\author{ Yves Brihaye\footnote{Yves.Brihaye@umh.ac.be}}
\address{Facult\'e des Sciences, Universit\'e de Mons-Hainaut,
 B-7000 Mons, Belgium}
\author{Betti Hartmann\footnote{Betti.Hartmann@durham.ac.uk}}
\address{Department of Mathematical Sciences, University
of Durham, Durham DH1 3LE, U.K.}
\date{\today}
\setlength{\footnotesep}{0.5\footnotesep}

\maketitle
\begin{abstract}

We study the direct interaction between global and local monopoles. While 
in two previous papers, the coupling between the two 
sectors was only ``indirect" through the coupling to gravity, we here 
introduce a new term in the potential that couples the Goldstone field
and the Higgs field directly. We investigate the influence of this
term in curved space and compare it to the results obtained
previously.

\end{abstract}


\renewcommand{\thefootnote}{\arabic{footnote}}

\section{Introduction}
Magnetic monopoles have raised a large interest since their
first construction by Dirac \cite{dirac}. While the Dirac-monopole
has a singularity, the so-called Dirac string, 't Hooft and Polyakov 
\cite{thooft} came up with the construction of a particle-like magnetic
monopole in SU(2) Yang-Mills-Higgs (YMH) theory with triplet Higgs scalar.
The magnetic charge of this object results from the topological
properties of the solution and is directly proportional
to the degree of the map from space-time infinite to the vacuum 
manifold of the theory. Minimal coupling of the SU(2) YMH model to
gravity leads (for suitable choice of the boundary conditions)
to globally regular gravitating monopoles \cite{nwp,weinberg,bfm,bizon}
which exist up to a maximal value of the gravitational coupling.
For higher values of that coupling, the Schwarzschild radius of the
solution becomes larger than the radius of the monopole core.

Considering only the theory with a scalar Goldstone field leads to
a different type of topological defect \cite{vilenkin}, the so-called global
monopole. Like all global defects this has infinite energy resulting
from the $\frac{1}{r^2}$ fall-off of the energy density. Coupling to
gravity \cite{vilenkin2,harari} leads to the observation that
the effective mass of the system becomes negative.
 
Recently, a self-gravitating magnetic monopole in the spacetime of
a global monopole has been considered \cite{spi,bh}. In both papers the
potential is the sum of 
the Higgs potential and the analog Goldstone field
potential. Thus, the interaction
between the global Goldstone field and the local Higgs field
is only indirect, namely through the coupling to gravity.

Considering the composite topological defect, 
the effective mass was found to be positive or negative, depending on 
the coupling constants of the model. 

In this paper we  continue the investigation of this system,
allowing a direct interaction between the matter fields. 
This extra interaction
is implemented by adding to the potential a gauge
invariant term as follows :
\begin{equation}
\label{pot}
V_3(\phi^a,\chi^a)=\frac{\lambda_3}2(\phi^a\phi^a-\eta_1^2)
(\chi^a\chi^a-\eta_2^2) \ .
\end{equation}

Here, we are mainly interested 
in the analysis of the critical behaviour of the
composite system considering now the most general, gauge invariant
potential. Because of the high non-linearity of the set of
coupled differential equation, an analytical 
analysis becomes impossible and only
a numerical analysis can provide the results.

This paper is organized as follows. In Section II we describe our model and 
the Ansatz. In Section III, we give the equations of motion, the boundary 
conditions and the analysis of the asymptotic behaviour of the Goldstone and 
Higgs field functions. We present our numerical 
results in Section IV and describe how the extra term (\ref{pot})  
in the potential 
provides new results concerning the behaviour of the fields near 
the defect's core as well as concerning the effective mass of the system. We  
observe e.g. a strong dependence of the mass on the coupling constant 
$\lambda_3$. We give our conclusions in Section V.

\section{The Extended Model}
This model is described by the following action which is 
composed of the action for the gravitating global monopole and the action
of the gravitating local monopole:
\begin{equation}
\label{action}
S=S_{G}+S_{M}=\int {\cal L}_{G}\sqrt{-g}d^{4}x+ \int {\cal L}_{M}\sqrt{-g}d^{4}x 
\ ,
\end{equation}
with the gravity Lagrangian ${\cal L}_G$:
\begin{equation}
{\cal L}_{G}=\frac{1}{16\pi G}R 
\end{equation}
and $G$ denotes Newton's constant.

The matter Lagrangian ${\cal L}_M$ of the extended model
with extra direct interaction between the global Goldstone field $\chi^a$ 
and the Higgs field $\phi^a$ reads ($a=1,2,3$):
\begin{equation}
\label{L}
{\cal L}_M=-\frac14 F_{\mu\nu}^a F^{\mu\nu,a}-\frac12(D_\mu\phi^a)
(D^\mu\phi^a)-\frac12(\partial_\mu \chi^a)(\partial^\mu\chi^a)-
V(\phi^a,\chi^a) \ , 
\end{equation}
with covariant derivative of the Higgs field 
\begin{equation}
D_\mu\phi^a=\partial_\mu\phi^a-e\epsilon_{abc}A_\mu^b\phi^c \ ,
\end{equation}
field strength tensor
\begin{equation}
F_{\mu\nu}^a=\partial_\mu A_\nu^a-\partial_\nu A_\nu^a-e\epsilon_{abc}
A_\mu^b A_\nu^c \ ,
\end{equation}
and $e$ being the gauge coupling constant.
The potential  $V(\phi^a,\chi^a)$ is given by:
\begin{eqnarray}
\label{coupling}
V(\phi^a,\chi^a)&=&\frac{\lambda_1}4\left(\phi^a\phi^a-\eta^2_1\right)^2+
\frac{\lambda_2}4\left(\chi^a\chi^a-\eta^2_2\right)^2\nonumber\\
&+&\frac{\lambda_3}2\left(\phi^a\phi^a-\eta_1^2\right)\left(\chi^a\chi^a-\eta^2_2
\right) \ ,
\end{eqnarray}
where the third term on the rhs couples the two sectors directly to each other
with coupling constant $\lambda_3$.
$\lambda_1$, $\lambda_2$ denote the self-coupling constants of the Higgs and
Goldstone field, respectively, while $\eta_1$, $\eta_2$ are the corresponding
vacuum expectation values.

The potential (\ref{coupling}) has different properties according to the 
sign of $\Delta \equiv  \lambda_1 \lambda_2 - \lambda_3^2$.
For $\Delta > 0$ the potential has positive values  and its minima are  
attained for  $\phi_a^2 = \eta_1^2$, $\chi_a^2 = \eta_2^2$,
for which (\ref{coupling}) is obviously zero.
For $\Delta <0$, these configurations  
become saddle points and two minima occur for
\be
\phi_a^2 = 0 \ , \ \ \chi_a^2 = \eta_2^2 + \frac{\lambda_1}{\lambda_3}\eta_1^2    
\ee
and
\be
\chi_a^2 = 0 \ , \ \ \phi_a^2 = \eta_2^2 + \frac{\lambda_2}{\lambda_3}\eta_2^2 \ .
\ee
The potential's values for these extrema are, respectively,
\be
         V_{min} = \frac{\eta_1^4}{4 \lambda_1} \Delta   \ \ , \ \
V_{min} = \frac{\eta_2^4}{4 \lambda_2} \Delta   
\ee
which are negative since $\Delta <0$.

\subsection{The Ansatz}
The Ansatz for the metric tensor in Schwarzschild-like coordinates reads:
\begin{equation}
ds^2=-A^2(r)N(r)dt^2+N^{-1}(r)dr^2+r^2(d\theta^2+\sin\theta^2d\phi^2) 
\end{equation}
where we define for later convenience the mass function $m(r)$ as follows~:
\begin{equation}
\label{N}
N(r)=1 -2 \alpha^2 q^2 -\frac{2m(r)}{r} \ . \ \
\end{equation}
The Ansatz for the global Goldstone field $\chi^a$, the Higgs field
$\phi^a$ and the gauge field $A_{\mu}^a$ in Cartesian coordinates reads:
\begin{equation}
\phi^a(x)=\eta_1 h(r)\hat{x}^a \ ,
\end{equation}
\begin{equation}
\chi^a(x)=\eta_1 f(r)\hat{x}^a \ ,
\end{equation}
\begin{equation}
A_i^a(x)=\epsilon_{iaj}\hat{x}^j\frac{1-u(r)}{er} \ ,
\end{equation}
and
\begin{equation}
\label{A0}
A_0^a(x)=0 \ .
\end{equation}
Substituting the above configurations into the matter Lagrangian density we 
obtain:
\begin{equation}
{\cal L}_M=-4\pi\int_0^\infty dr \ r^2 \ A\left[N {\cal{K}}(f,h,u)+{\cal{U}}(f,h,u)
\right] \ ,
\end{equation}
where
\begin{equation}
\label{K}
{\cal K}(f,h,u)=\frac12\eta^2_1(f')^2+\frac12\eta^2_1(h')^2+\frac{(u')^2}
{e^2r^2} \ ,
\end{equation}
and
\begin{eqnarray}
{\cal U}(f,h,u)&=&\frac{(u^2-1)^2}{2e^2r^4}+\frac{\eta^2_1u^2 h^2}{r^2}+
\frac{\eta^2_1f^2}{r^2}+\frac{\lambda_1\eta^4_1}4(h^2-1)^2+
\frac{\lambda_2\eta^4_1}4(f^2-q^2)^2\nonumber\\
&+&\frac{\lambda_3\eta_1^4}2(h^2-1)(f^2-q^2) \ .
\end{eqnarray}
The prime  denotes the derivative with respect to $r$. 

The gravity Lagrangian ${\cal L}_G$ is given 
by
\begin{equation}
{\cal L}_G=\frac1{2G}\int_0^\infty dr r (N-1)A' \ .
\end{equation}

\section{Equations of Motion}

Varying (\ref{action}) with respect to the matter fields and gravitational
fields and introducing the dimensionless variable $x$ and dimensionless mass function
$\mu(x)$ :
\begin{equation}
x=e\eta_1 r \ , \ \mu(x)=e\eta_1 m(r) 
\end{equation} 
we obtain the following set of differential equations: 
\begin{equation}
\frac d{dx}\left[x^2AN\frac {df}{dx}\right]=
A\left[2f+x^2\beta_2^2f(f^2-q^2)+
{x^2\beta_3^2} (h^2-1)f\right] \ ,
\end{equation}
\begin{equation}
\frac d{dx}\left[ x^2AN\frac {dh}{dx}\right]=A\left[2u^2h+x^2\beta_1^2(h^2-1)h
+{x^2\beta_3^2}h(f^2-q^2)\right] \ ,
\end{equation}
\begin{equation}
\frac d{dx}\left[AN\frac{du}{dx}\right]=A\left[\frac{u(u^2-1)}{x^2}+
uh^2\right] \ ,
\end{equation}
\begin{equation}
\label{xAN}
\frac d{dx}(xAN)=[1-2\alpha^2x^2\bar{U}]A \ ,
\end{equation}
with
\begin{eqnarray}
\bar{U}&=&\frac{(u^2-1)^2}{2x^4}+\frac{u^2h^2}{x^2}+\frac{f^2}{x^2}+
\frac{\beta_1^2}4(h^2-1)^2+\frac{\beta_2^2}4(f^2-q^2)^2\nonumber\\
&+&\frac{\beta_3^2}2(h^2-1)(f^2-q^2) \ ,
\end{eqnarray}
and
\begin{equation}
\frac{dA}{dx}=2\alpha^2 Ax\bar{K}
\end{equation}
with
\begin{equation}
\bar{K}=\frac12\left(\frac{df}{dx}\right)^2+\frac12\left(\frac{dh}{dx}
\right)^2+\frac1{x^2}\left(\frac{du}{dx}\right)^2 \ .
\end{equation}
The equations only depend on the dimensionless coupling constants:
\begin{equation}
\alpha^2=4\pi G\eta_1^2 \ , \ \ \  \beta_i^2=\lambda_i/e^2 \ , \
i=1, 2, 3 \ , \ \ \  q=\eta_2/\eta_1 \ .
\end{equation}
With the definition (\ref{N}), Eq. (\ref{xAN}) can be brought to the form
\begin{equation}
\mu' = \alpha^2 x^2 (\bar K + (\bar U - \frac{q^2}{x^2})) \ .
\label{finiteeng}
\end{equation}
The {\it finite} energy of the solution can then be
obtained by taking the value of $\mu(x)$ at infinity.

\subsection{Boundary conditions}
The boundary conditions at the origin which follow from the
requirement of
regularity read:
\begin{equation}
\label{O}
u(x=0)=1 \ , \ \ f(x=0)=0\ , \ \ h(x=0)= 0 \ , \ \ \mu(x=0)=0.
\end{equation}
In fact the behaviour of the function $\mu(x)$ near the origin is $\mu(x)
\approx -\alpha^2 q^2 x$. The requirement of finite energy solutions
leads to:
\begin{equation}
f(x=\infty)=  q \ , \ \ h (x=\infty)=  1 \ , \ \ u(x=\infty)= 0 \ , \ \ N
(x=\infty)=1-2\alpha^2 q^2.
\end{equation} 
\subsection{Asymptotic behaviour}

The integration of the equations for generic values of the parameters
needs a better understanding of the asymptotic behavior of the
functions $f(x)$ and $h(x)$. Two different types of behaviour for the functions
$f$ and $h$ in flat space seem possible. Either:
\begin{eqnarray}
\label{podecay}
&h(x) = 1 + \frac{A}{x^2} + O(\frac{1}{x^3}) \ \ \ , \ \ 
&A = \frac{\beta_3^2}{\beta_2^2 \beta_1^2 - \beta_3^4} \nonumber \\
&f(x) = q + \frac{B}{x^2} + O(\frac{1}{x^3}) \ \ \ , \ \ 
&B = \frac{- \beta_1^2}{q(\beta_2^2 \beta_1^2 - \beta_3^4)} 
\end{eqnarray}
or 
\begin{eqnarray}
\label{exdecay}
&h(x) &= 1 + C_1 \exp(\rho_1 x) + C_2 \exp(-\rho_1 x) + 
C_3 \exp(\rho_2 x) + C_4 \exp(-\rho_2 x)
 , \ \nonumber \\ 
&f(x) &=q + \tilde C_1 \exp(\rho_1 x) + \tilde C_2 \exp(-\rho_1 x) +
\tilde C_3 \exp(\rho_2 x)  + \tilde C_4 \exp(-\rho_2 x)
 \ , \
\end{eqnarray}
where $\rho_1^2$, $\rho_2^2$ are the eigenvalues of the matrix
\be
\label{matrix}
      \left(\matrix{\beta_1^2 &q \beta_3^2\cr
                    q \beta_3^2 & q^2 \beta_2^2\cr}\right) \ .
\ee
Our numerical analysis strongly suggests the following results~: 
\begin{itemize}
\item{}  for $\beta_1^2 \beta_2^2 - \beta_3^4 > 0$ the solutions obey the 
asymptotic behaviour (\ref{podecay}) for $f$ and $h$,

\item{}  for $\beta_1^2 \beta_2^2 - \beta_3^4 < 0$
the  functions $f$ and $h$ have the  
asymptotic behaviour (\ref{exdecay}). In this case, however, one of the
eigenvalues of (\ref{matrix}) becomes negative, 
consequently its square root 
is complex and the functions $f,h$ thus oscillate. This behaviour is clearly 
observed when we set $\beta_1=0$, $\beta_2 \neq 0$ and increase $\beta_3$
from 0. For $\beta_3=0$, the solutions exist and $f,h$ increase monotonically, 
as soon as $\beta_3 \neq 0$, however, oscillations occur.
\end{itemize}
\section{Numerical results}

Because of the reasons given previously, we restrict our analysis in the following 
to the case $\beta_1^2 \beta_2^2 - \beta_3^4 > 0$.

The limit $\beta_3=0$ was studied in detail in \cite{spi}, 
\cite{bh}. Here we discuss how the new term influences the 
behaviour of the solutions.
One of the main features is that the Higgs function $h(x)$ does
not reach its asymptotic value $h(x=\infty)=1$ monotonically. It first 
reaches
a maximum  $h_{max} > 1$ for $x < \infty$  and then decreases to 1.
This phenomenon, which can be expected from the inspection
of (\ref{podecay}) since $A >0$ for $\beta_3\neq 0$, 
is illustrated in Fig.~1.
This is different from the phenomena observed in \cite{spi}, \cite{bh}.
There, for all values of the coupling constants, the function $h(x)$ was
observed to be monotonically increasing from 0 to 1 as indicated in Fig. 
1 for $\beta_3=0$. The Goldstone field functions $f(x)$
reaches its asymptotic value $q$ for increasing values of the
coordinate $x$ when $\beta_3$ is increasing. This again can be explained
by (\ref{podecay}) since the value $B$ is a decreasing function of $\beta_3$
with a sharp drop at $\beta_3\approx 1$. Thus, the function $f(x)$ decays less
strong for higher values of $\beta_3$.  

At the same time, the minimum of the
metric function $N$ decreases with increasing $\beta_3$, while
the matter function $\mu$ reaches a maximum at roughly the same
$x$ at which the function $N$ attains its minimum and than
drops down to its asymptotic value which determines the
mass of the solution. Clearly, for $\beta_3=0$, this asymptotic value
is positive, while for increasing $\beta_3$, it decreases and
becomes negative for large enough $\beta_3$.

Another feature of the
new term is that, for all parameters but $\beta_3$ fixed, the
classical mass of the solution decreases when $\beta_3$ increases.
This is illustrated by means of Fig.~2 where we have plotted the 
evolution of the mass as a function
of $\beta_3$ for three different combinations of the coupling
constants $\alpha,q$.
For $q=1.0$, $\alpha=0.4$, the mass of the solution is already negative
for $\beta_3=0$ indicating that the influence of the global monopole is 
already dominating in the limit of vanishing direct interaction of the
Higgs field and Goldstone field. For smaller $q$, the mass becomes 
negative at some finite value $\beta_3=\beta_3^0$. For fixed $q$, this 
value is increasing for decreasing $\alpha$. Since the local and global 
monopole are still ``indirectly'' coupled over gravity, a stronger 
gravitational coupling, of course, couples the two objects in a stronger
way. For small gravitational coupling, $\beta_3$ thus has to be raised 
further to make the influence of the global monopole dominating. When the 
combination $\beta_1^2 \beta_2^2-\beta_3^4$ becomes negative, the mass
of the solution reaches $-\infty$, independent of the combination
of $q$ and $\alpha$. This again can be related to the fact that we observe
the solutions to become oscillating for $\beta_1^2 \beta_2^2-\beta_3^4 < 
0$. 

We also studied the way the solution bifurcates into a black hole
when the parameter $\alpha$ increases while the others are fixed.
Fixing $q=0.4$ we have analyzed the critical behaviour of the solution
and checked that, like for the case $\beta_3=0$, the solution bifurcates
into a black hole for a finite value of $\alpha$, say $\alpha=\alpha_c$.
As demonstrated in Fig.~3 for $\beta_1=\beta_2=1$, $q=0.4$ and
$\beta_3=0.8$, the 
function $N$ develops a minimum which 
becomes
deeper while $\alpha$ increases and becomes zero for $\alpha = \alpha_c$.
The limiting solution thus represents an extremal black hole solution
with horizon $x_h$. In Fig.~3, we also show the evolution of the mass
with $\alpha$.
Finally, the critical solution corresponding to $\beta_3=0.8$ 
and $\alpha_c \approx 1.085$
is  displayed in Fig.~4.
This figure clearly suggests that the limiting solution is not an abelian
black hole for $x > x_h$ like e.g. in the case of the pure local monopole
\cite{bfm}, where the solutions bifurcate with the branch
of Reissner-Nordstr\"om (RN) solutions and consequently
the functions reach their RN values for $x > x_h$.  
Here, the function $f$ is equal to zero for
$0\leq x\leq x_h$ and non-trivial for $x_h\leq x\leq x_0$. Similarly,
the functions $A$ and $h$ are non-trivial for $x\leq x_0$ and are
equal to their asymptotic values for $x > x_0$, while the gauge
field function $u$ reaches its asymptotic value
for $x\approx x_h$. This solution thus represents a ``black hole inside a global 
monopole'' as was observed previously for the $\beta_3=0$ limit \cite{bh}.

\section{Conclusions}

In this paper we have analyzed the composite system 
of a global and a local gravitating monopole
considering the most general gauge-invariant potential. 
This potential contains a direct interaction between the Golstone
and the Higgs field. This term leads to
important consequences concerning the local behaviour of the fields 
themselves as well as concerning the global properties of the system. One of the most
relevant consequences is related to the effective mass associated with the
composite topological defect. The numerical results show a strong dependence
of this mass on the coupling constant $\lambda_3$. Although the
Goldstone and Higgs fields are indirectly coupled through
gravity, the extra 
direct interaction is more effective. Increasing the parameter $\lambda_3$, 
the mass
becomes negative, indicating the dominance of the global sector
over the local one. Compared to the results of the $\lambda_3=0$ case
\cite{spi,bh}, we observe the modulus of the negative mass
to become very large in our system.
Another point which deserves to be mentioned is that the extra direct 
interaction term is not positive definite. 
Denoting by $x_{h=1}$ the value of $x$ for which the Higgs field function
$h$ is equal to one, $h$ becomes bigger than one for $x >x_{h=1}$
and the new term in the potential becomes 
negative. However, for a particular choice of the self-coupling constants
such that they fulfill $\Delta\equiv \lambda_1\lambda_2-\lambda_3^2>0$,
the total potential is positive, vanishing
only at the minima $\phi_a^2=\eta_1^2$, $\chi_a^2=\eta_2^2$. 

As possible extensions of the model studied here, let us mention
the coupling to a scalar dilaton which arises naturally in low energy effective
actions of string theory. The gravitating local monopole was studied recently
coupled to a dilaton \cite{bhk} and it was found that in the limit
of critical gravitational coupling, the solutions bifurcate with the branch
of extremal Einstein-Maxwell-dilaton (EMD) solutions which are associated
with naked singularities. It would be interesting to see what sort of
critical solution the
composite system of a global and local monopole reaches since our analysis
indicates that the behaviour of the functions
close to the core of the local monopole
is strongly influenced by the global monopole. 

\newpage
\begin{figure}\centering\epsfysize=20cm
\mbox{\epsffile{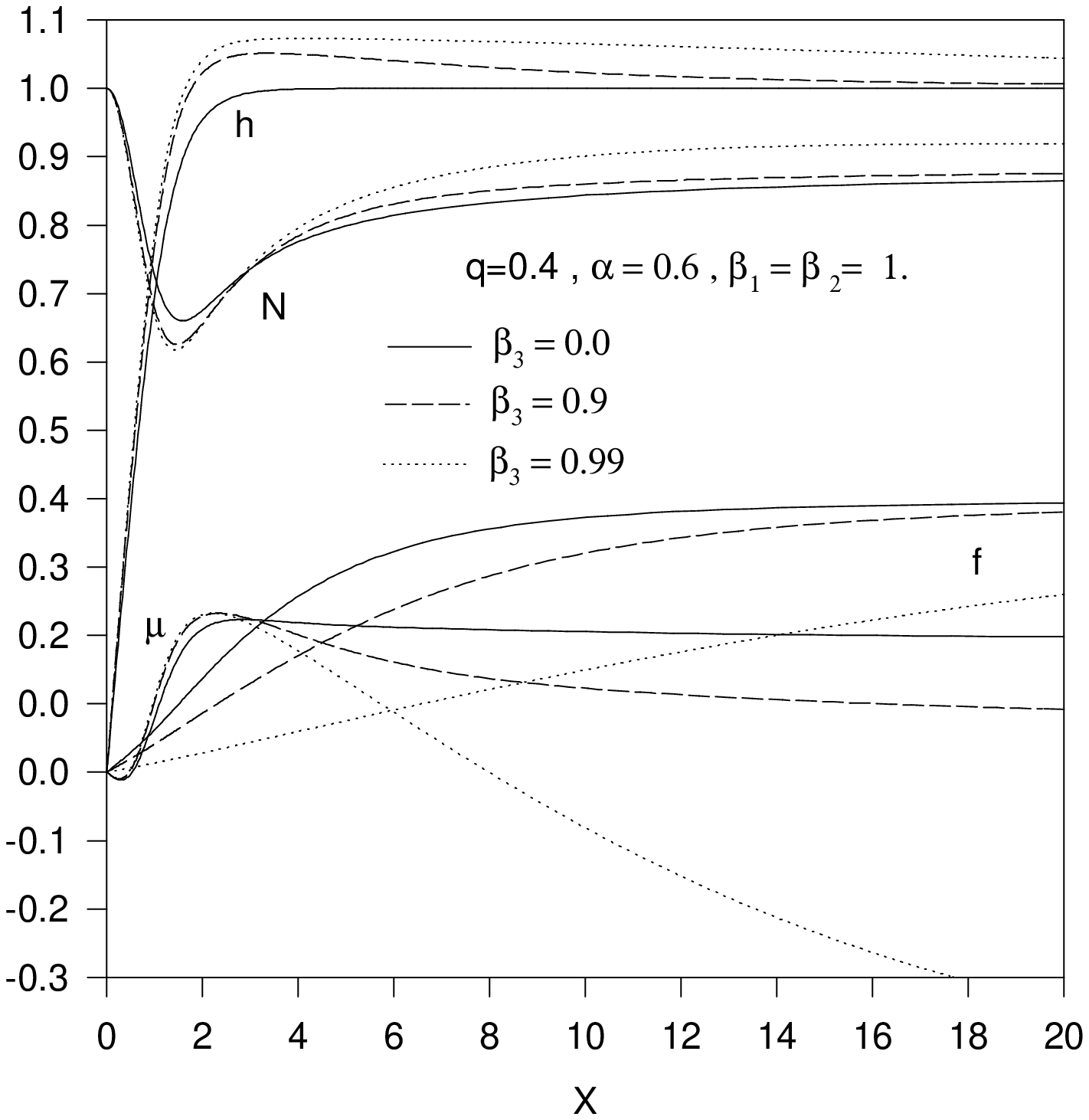}}
\caption{The metric functions $N$, $\mu$, the Higgs field function $h$ and the Goldstone
field function $f$ are shown
as functions of the dimensionless variable $x$ for $q=0.4$, $\alpha=0.6$, $\beta_1=\beta_2=1$
and three different values of $\beta_3$. }
\end{figure}
\begin{figure}\centering\epsfysize=20cm
\mbox{\epsffile{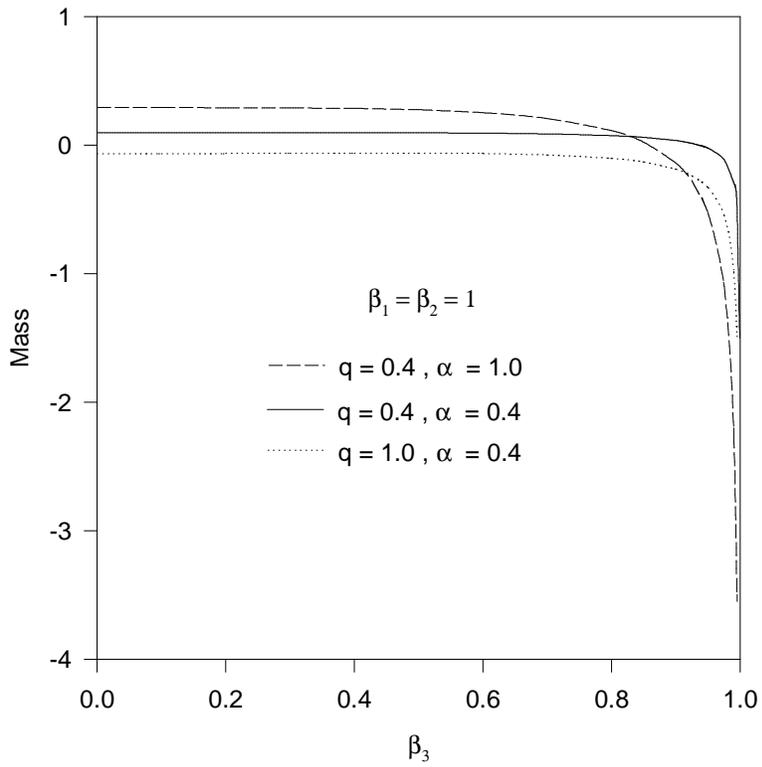}}
\caption{The mass in units of $\frac{4\pi \eta_1}{e}$ is shown as  
function
of $\beta_3$ for three different combinations  of the coupling constants 
$\alpha$ and $q$ and for $\beta_1=\beta_2 =1$.}
\end{figure}
\begin{figure}\centering\epsfysize=20cm
\mbox{\epsffile{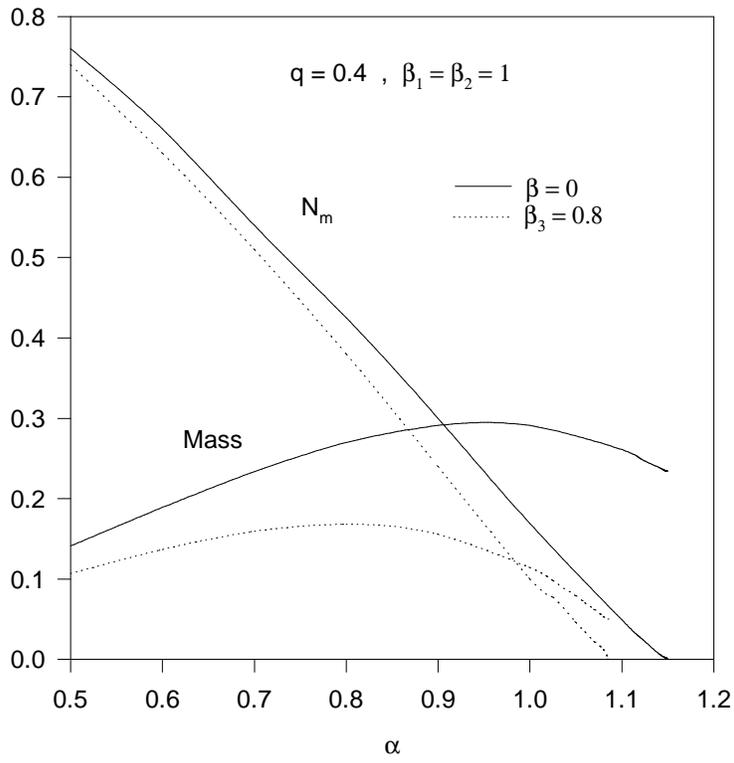}}
\caption{The mass in units of $\frac{4\pi\eta_1}{e}$ and the minimum
$N_m$ of the metric function $N(x)$ are shown as function of $\alpha$ for
$\beta_3=0$, respectively $\beta_3=0.8$, and $q=0.4$, $\beta_1=\beta_2=1$.} 
\end{figure}
\begin{figure}\centering\epsfysize=20cm
\mbox{\epsffile{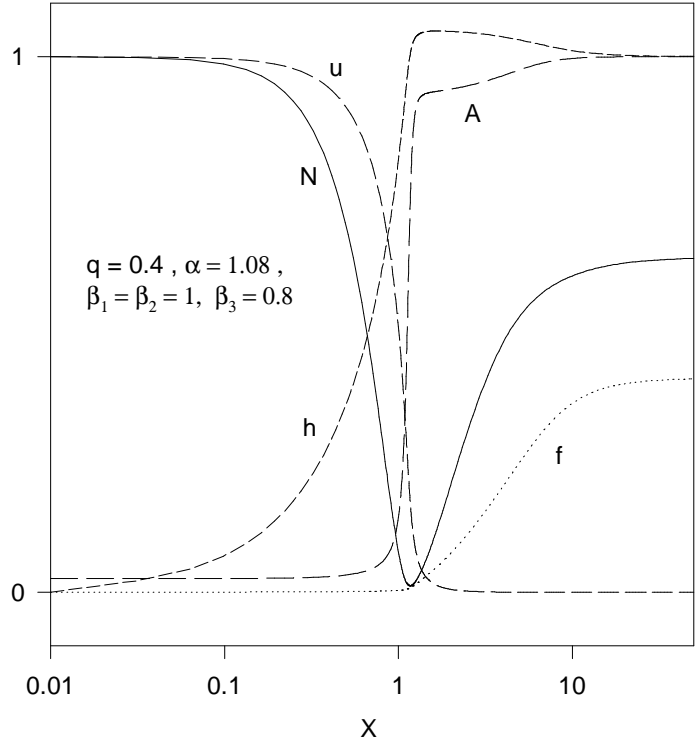}}
\caption{The profiles of the functions $N,A, u,f,h$ are shown for 
$\alpha\approx\alpha_c=1.08$, $\beta_1 = \beta_2 = 1$, $\beta_3=0.8$
and $q=0.4$.} 
\end{figure}
\end{document}